\documentclass[a4paper,UKenglish,cleveref, autoref]{lipics-v2019}

\usepackage[utf8]{inputenc}
\usepackage{textgreek}
\usepackage{amssymb}
\usepackage{graphicx}
\usepackage{url}
\usepackage{alltt}
\usepackage{amsmath}
\usepackage{proof}
\usepackage{todonotes}
\usepackage{newunicodechar}

\newunicodechar{◂}{$\blacktriangleleft$}
\newunicodechar{★}{$\star$}
\newunicodechar{➔}{$\rightarrow$}
\newunicodechar{∀}{$\forall$}
\newunicodechar{ᵢ}{$_i$}
\newunicodechar{₁}{$_1$}
\newunicodechar{₂}{$_2$}
\newunicodechar{≃}{$\simeq$}
\newunicodechar{∃}{$\exists$}
\newunicodechar{➾}{$\Rightarrow$}
\newunicodechar{≤}{$\leq$}
\newunicodechar{·}{$\!\!\!$}

\newcommand{\authors}{Denis Firsov, Larry Diehl, Christopher Jenkins, and Aaron Stump}

\title{Course-of-Value Induction in Cedille} 

\titlerunning{}

\author{\authors}{University of Iowa, USA}{\{firstname-lastname\}@uiowa.edu}{}{}

\authorrunning{D. Firsov, L. Diehl, C. Jenkins, and A. Stump}

\Copyright{\authors}



\category{}

\relatedversion{}

\supplement{}

\nolinenumbers 

\EventEditors{John Q. Open and Joan R. Access}
\EventNoEds{2}
\EventLongTitle{International Conference on Formal Structures for Computation and Deduction (FSCD 2019)}
\EventShortTitle{FSCD 2019}
\EventAcronym{FSCD}
\EventYear{2019}
\EventDate{June 24--30, 2019}
\EventLocation{Dortmund, Germany}
\EventLogo{}
\SeriesVolume{42}
\ArticleNo{23}

\begin{document}

\maketitle

\begin{abstract}
In the categorical setting, histomorphisms model a course-of-value
recursion scheme that allows functions to be defined using arbitrary
previously computed values. In this paper, we use the Calculus of
Dependent Lambda Eliminations (CDLE) to derive a lambda-encoding of
inductive datatypes that admits course-of-value induction. Similar to
course-of-value recursion, course-of-value induction gives access to
inductive hypotheses at arbitrary depth of the inductive arguments of
a function. We show that the derived course-of-value datatypes are
well-behaved by proving Lambek's lemma and characterizing the
computational behavior of the induction principle. Our work is
formalized in the Cedille programming language and also includes
several examples of course-of-value functions.
\end{abstract}

\section{Introduction}
\label{intro}
Dependently typed programming languages with built-in infrastructure
for defining inductive datatypes allow programmers to write functions with
complex recursion patterns. For example, in Agda~\cite{lang:agda} we can implement
the natural definition of Fibonacci numbers:
\begin{alltt}
fib : Nat → Nat
fib zero = zero
fib (suc zero) = 1
fib (suc (suc n)) = fib (suc n) + fib n
\end{alltt}
This definition is accepted by Agda because its built-in
termination checker sees that all recursive calls are done on
structurally smaller arguments.
In contrast, in pure polymorphic lambda calculi (e.g., System F),
inductive datatypes can be encoded by means of impredicative
quantification (without requiring additional infrastructure). For
example, if we assume that \verb;F; is a well-behaved positive scheme
(e.g., a functor), then we can express its least fixed
point as an initial Mendler-style F-algebra. A Mendler-style algebra
differs from a traditional F-algebra (\verb;F X → X;) in that it takes
an additional argument (of type \verb;R → X;), which corresponds to a
function for making recursive calls.
Mendler-style algebras introduce a polymorphic type \verb;R; for
\textit{recursive} scheme arguments, allowing recursive function calls
to be restricted to structurally smaller arguments.
At the same time, the polymorphic type prevents any kind of further
inspection of those arguments (for the remainder of this paper we
switch to code written in Cedille~\cite{lang:cedille}, a dependently
typed programming language supporting impredicative type
quantification):
\begin{alltt}
AlgM ◂ (★ ➔ ★) ➔ ★ ➔ ★ = λ F. λ X. ∀ R : ★. (R ➔ X) ➔ F · R ➔ X.
FixM ◂ (★ ➔ ★) ➔ ★ = λ F. ∀ X : ★. AlgM F X ➔ X.
foldM ◂ ∀ F : ★ ➔ ★. ∀ X : ★. AlgM F X ➔ FixM F ➔ X
 = Λ F. Λ X. λ alg. λ v. v alg.
\end{alltt}
The simple recursion pattern provided by \verb;foldM; (also known as
catamorphism) can be tricky to work with. Let us define natural
numbers as the least fixed point of the functor \verb;NF X = 1 + X;.
Hutton~\cite{hutton99} explained that it is possible to use the
universality law of initial F-algebras to show that there is no
algebra \verb;g : AlgM NF Nat; such that \verb;fib = foldM g;. The
reason for this is that in the third equation (of the natural
definition of \verb;fib;), the recursive call is made not only on the
direct predecessor of the argument (\verb;suc n;), but also on the
predecessor of the predecessor (\verb;n;).

The usual workaround involves ``tupling''. More specifically, we
define an algebra \verb;AlgM NF (Nat × Nat); where the second \verb;Nat;
denotes the previous Fibonacci number. Then, we fold the input with
\verb;fibAlg;, and finally return the first projection of the tuple
(here \verb;πᵢ; denotes \verb;i;-th projection from the tuple):
\begin{alltt}
fibAlg ◂ AlgM NF (Nat × Nat) = Λ R. λ rec. λ fr. 
 case fr (λ _.  pair (suc zero) zero)      % zero case
         (λ r. let p = rec r in            % suc case
               pair (add (π₁ p) (π₂ p)) (π₁ p)).
fibTup ◂ Nat ➔ Nat = λ n. π₁ (foldM fibAlg n).
\end{alltt}
In this example, the \verb;rec; function allows recursive calls to be
made explicitly on elements of type \verb;R; (which is \verb;Nat; in
disguise). This approach requires error-prone bookkeeping.
Additionally, observe that the defining equation of the Fibonacci numbers
(\texttt{fibTup (suc (suc n)) = fibTup n + fibTup (suc n)}) is true
propositionally, but not definitionally (i.e., it does not follow by
$\beta$-reduction).

The alternative solution to tupling is course-of-value recursion
(also known as histomorphism), which makes it possible to express
nested recursive calls directly. The central concept of this approach is
course-of-value F-algebras, which are similar to usual Mendler-style
F-algebras, except that they take an abstract destructor (of type
\verb;R ➔ F R;) as yet another additional argument. The abstract
destructor is a fixed-point unrolling (or, abstract inverse of the
initial algebra), and intuitively allows for ``pattern-matching'' on
constructors for the scheme \verb;F;.
\begin{alltt}
AlgCV ◂ (★ ➔ ★) ➔ ★ ➔ ★ = λ F. λ X.
  ∀ R : ★. (R ➔ F · R) ➔ (R ➔ X) ➔ F · R ➔ X. 
\end{alltt}
For illustration purposes, assume that \verb;F; is a functor and that
\verb;FixCV F; is its least fixed point. Also, assume that \verb;inCV;
and \verb;outCV; are mutual inverses and 
represent a collection of constructors and destructors, respectively.
\begin{alltt}
inCV  ◂ F (FixCV F) ➔ FixCV F = <..>
outCV ◂ FixCV F ➔ F (FixCV F) = <..>
\end{alltt}
Then, course-of-value recursion is characterized by the function
\verb;foldCV;, and its reduction behaviour is characterized by the
cancellation property (\verb;cancel;):
\begin{alltt}
foldCV ◂ ∀ X : ★. AlgCV F X ➔ FixCV F ➔ X = <..>
cancel ◂ ∀ X : ★. ∀ alg : AlgCV F X. ∀ fx : F (FixCV F).
  foldCV alg (inCV fx) ≃ alg outCV (foldCV alg) fx = β.
\end{alltt}
Notice that after unfolding, the first argument of \verb;alg; is
instantiated with \verb;outCV; (the destructor), and the second
argument is instantiated with a partially applied \verb;foldCV; (the
recursive call).

To illustrate the nested recursive calls, we define course-of-value
naturals (\verb;NatCV;) as the course-of-value least fixed point of the
functor \verb;NF; (\verb;FixCV NF;). Then, the Fibonacci function can
be implemented very close to the conventional ``pattern-matching'' style:
\begin{alltt}
fibCV ◂ NatCV ➔ NatCV = foldCV (Λ R. λ out. λ rec. λ nf. 
 case nf (λ _. zero)                       % zero case
  (λ r. case (out r) (λ _. suc zero)       % (suc zero) case
             (λ r'. rec r + rec r'))).     % (suc (suc n)) case
\end{alltt}
Here, \verb;out; provides an additional layer of pattern-matching on the
arguments of the function.  Finally, it is important to observe that
given that \verb;cancel; is true by β-reduction, then
\verb;fibCV (suc (suc n)) ≃ fibCV (suc n) + fibCV n; is also true by
β-reduction.

The remaining questions is how to define the least fixed point
\verb;FixCV F; for any positive scheme \verb;F;, and how to derive the
corresponding introduction and elimination principles. We might try a
usual construction in terms of universal quantification and
\verb;AlgCV;:
\begin{alltt}
FixCV ◂ (★ ➔ ★) ➔ ★ = λ F. ∀ X : ★. AlgCV F X ➔ X.
\end{alltt}
This definition fails since \verb;AlgCV F X; is
isomorphic to \verb;AlgM (Enr' F) X; where 
\begin{alltt}
Enr' ◂ (★ ➔ ★) ➔ ★ ➔ ★ = λ F. λ X. F X × (X ➔ F X).
\end{alltt}
\verb;Enr' F; is a \textit{negative} scheme and (in general) the
least fixed points of negative schemes are undefined in a consistent
type theory.  As a result, it is common to implement \verb;foldCV; in
terms of general recursion from the host language or to add it as a
primitive construction (see related work in Section~\ref{related-work}).

The \textbf{main contribution} of this paper is the derivation of
course-of-value datatypes in the Calculus of Dependent Lambda
Eliminations (CDLE). The key inspiration for our work comes from the
categorical construction known as \emph{restricted existentials}
(Section~\ref{restricted-existentials}). We prove that the least fixed
point of the restricted existential of scheme \verb;Enr' F; exists,
and that it contains course-of-value datatypes as a subset
(Section~\ref{precursor}). Next, we employ heterogeneous equality
from CDLE to define the datatype \verb;FixCV; as this subset
(Section~\ref{cov-induction}).  We also give (according to our best
knowledge) the first generic formulation and derivation of
a course-of-value induction principle in a pure type theory.  Finally, we
show examples of functions and proofs defined over course-of-value natural
numbers (Section~\ref{examples}).

The CDLE type theory is implemented in Cedille, which we use to
type-check the formalized development of this paper.\footnote{
  The Cedille formalization accompanying this paper is available at:\\
  \url{http://firsov.ee/cov-induction}
} We emphasize that part of the significance of our work is
\textit{deriving} course-of-value induction within the small core calculus of
CDLE. While the Calculus of Inductive Constructions (CIC)~\cite{coq}
directly adds support for inductive datatypes to the Calculus of Constructions (CC)
in an involved way, CDLE is a minor extension of CC that makes inductive
datatypes derivable rather than built-in.

\section{Background}

\subsection{The CDLE Type Theory}
\label{cedille}

CDLE~\cite{stump17a,stump18} is an extrinsically typed (or,
Curry-style) version of the Calculus of Constructions (CC), extended
with a heterogeneous equality type (\verb;t₁ ≃ t₂;)\footnote{The most
  recent version of CDLE~\cite{lang:cedille} has been extended with a
  more expressive equality type, but this work does not make use of
  it.
}, Kopylov's~\cite{kopylov03} dependent intersection type
(\verb;ι x : T. T';), and Miquel's~\cite{miquel01} implicit product
type (\verb;∀ x : T. T';, for erased arguments).
To make type checking algorithmic, Cedille
terms have typing annotations, and definitional equality of terms is
modulo erasure of these annotations. The target of erasure in Cedille
is the pure untyped lamba calculus with no additional constructs. Due
to space constraints, we omit a more detailed summary of CDLE.
However, this work is a direct continuation of our previous
work~\cite{firsov18b}, which includes a detailed explanation of
all of the constructs of Cedille.

\subsection{Identity Functions and Identity Mappings}
Stump showed how to derive induction for natural numbers in
CDLE~\cite{stump17b}. This result was generically extended to
achieve induction for arbitrary datatypes that arise as a least fixed point of
a functor~\cite{firsov18}.
One final generalization restricted \verb;fmap; to be defined over
functions that extensionally behave like identity functions (the type
\verb;Id;). Least fixed points of such ``identity mapping'' (the type
\verb;IdMapping;) schemes can be defined for a larger collection of datatypes,
compared to functors. We omit the implementations below
(indicated by \verb;<..>;), but a detailed description of these
constructs can be found in~\cite{firsov18b}.

\label{identity-functions}
\paragraph{\textbf{Identity Functions}} We define the type \verb;Id X Y; as the
collection of all functions from \verb;X; to \verb;Y; that erase to the term
(\verb;λ x. x;):
\begin{alltt}
Id ◂ ★ ➔ ★ ➔ ★ = λ X. λ Y. Σ f : X ➔ Y. f ≃ (λ x. x).
\end{alltt}
Because Cedille is extrinsically typed, the domain (\verb;X;)
and codomain (\verb;Y;) of an identity function need not be the same.
An identity function \verb;Id X Y; can be eliminated to ``cast''
values of type \verb;X; to values of type \verb;Y; without changing
the values themselves:
\begin{alltt}
elimId ◂ ∀ X Y : ★. ∀ c : Id · X · Y. X → Y = <..>
\end{alltt}

Note that argument \verb;c; of \verb;elimId; is
quantified using \verb;∀; (rather than \verb;Π;), indicating that it is
an implicit (or, erased) argument. Significantly, the erasure
\verb;|elimId -c|; (the dash syntactically
indicates that this an implicit, or erased, application)
results in the identity function (\verb;λ x. x;)\footnote{
  Types, as opposed to values, are always erased in terms. Hence,
  using (\texttt{∀ X : ★}) in the classifier of a term is sensible but
  using (\texttt{Π X : ★}) is not. Additionally, we omit type
  applications in terms because they are inferred by Cedille.}.

\paragraph{\textbf{Identity Mappings}} We say that scheme \verb;F; is an identity mapping
if it is equipped with a function that lifts identity functions:
\label{identity-mapping}
\begin{alltt}
IdMapping ◂ (★ ➔ ★) ➔ ★ = λ F. ∀ X Y: ★. Id X Y ➔ Id (F X) (F Y).
\end{alltt}
Intuitively, \verb;IdMapping F; is
similar to a functor's \verb;fmap;, but it only needs to be defined on
identity functions, and no additional laws are required.
Every functor induces an identity mapping, but not vice
versa~\cite{firsov18b}.

\subsection{Inductive Datatypes in Cedille}
\label{inductive-datatypes}
Next, we review the generic datatype constructions (from \cite{firsov18b})
definable using the constructions above.
To start with, we specify a scheme \verb;F; and its identity mapping as
module-level parameters (such that every definition in this section
begins with these parameters).
\begin{alltt}
module (F : ★ ➔ ★)\{imap : IdMapping · F\}.
\end{alltt}
Curly braces around the \verb;imap; variable indicate that it is
quantified implicitly (or, as an erased parameter). Another way of
saying this is that none of the definitions should depend on the
computational behaviour of \verb;imap;.

The fixpoint type (\verb;FixIndM ◂ ★;) is defined as an
intersection of \verb;FixM; and a proof of its inductivity
(see~\cite{firsov18b} for details). \verb;FixIndM; comes with
a constructor (\texttt{inFixIndM  ◂ F · FixIndM ➔ FixIndM}),
and its mutually inverse
destructor (\texttt{outFixIndM ◂ FixIndM ➔ F · FixIndM}).
The induction principle for \verb;FixIndM; takes a
``dependent'' counterpart to Mendler-style F-algebras (\verb;AlgM;),
which we call Q-proof-algebras (\verb;PrfAlgM;):
\begin{alltt}
induction ◂ ∀ Q : FixIndM ➔ ★. PrfAlgM Q ➔ Π e : FixIndM. Q e = <..>
\end{alltt}
A value of type \verb;PrfAlgM Q; should be understood as an inductive
proof that predicate \verb;Q; holds for every \verb;FixIndM; built by
constructors \verb;inFixIndM;.
Just like F-algebras, Q-proof-algebras allow users to invoke inductive
hypotheses only on direct subdata of a given argument. The rest of the
paper is devoted to the formulation and derivation of a generic
course-of-value induction principle that allows users to invoke
inductive hypotheses on subdata at arbitrary depths
(realized as \verb;inductionCV; for any \verb;PrfAlgCV;
in Section~\ref{cov-induction}).

\section{Restricted Existentials}
\label{restricted-existentials}
Uustalu and Vene defined a construction called the restricted
existential to demonstrate an isomorphism between Church-style and
Mendler-style initial algebras~\cite{uustalu99}. The importance of
this is that for any difunctor (or, mixed variant functorial scheme)
\verb;F;, the restricted existential of \verb;F; is an isomorphic
covariant functor.

In this section, we define a variation that we call an \emph{identity
  restricted existential}.  We also derive its dependent elimination
principle, and prove that the identity restricted existential of any
scheme \verb;F; (including negative and non-functorial ones) is an
\emph{identity mapping}. Later in the paper, the restricted
existential will be the main tool for deriving course-of-value
datatypes.

\subsection{Restricted Coends}
In the categorical setting, the restricted existential arises as a
restricted coend. Our subsequent development requires
existentials where the quantifier ranges over types. This can be
provided by a restricted coend (\verb;RCoend H F;), which is
isomorphic to the existential type \verb;∃ R. H R × F R; (where
\verb;H; is what we are restricting by). Our development defines
\verb;RCoend; by taking advantage of the isomorphism between the
universal type \verb;∀ R. H R ➔ F R ➔ Q; and the existential type
\verb;(∃ R. H R × F R) ➔ Q; (for any \verb;Q;) that we have in
mind. Now, let us formalize the notion of restricted coend.

Let $F : \mathcal{C}^{op} \times \mathcal{C} \rightarrow \mathcal{C}$ be an
endodifunctor and $H : \mathcal{C}^{op} \times \mathcal{C} \rightarrow \mathit{Set}$
be a difunctor to $\mathit{Set}$. An $H$-restricted $F$-coend is an initial
object in the category of $H$-restricted $F$-cowedges. An $H$-restricted
$F$-cowedge is a pair $(C, \Phi)$ where $C$ (the carrier) is an object in
$\mathcal{C}$ and $\{\Phi_R\}_{R \in \mathcal{C}}$ is a family of
functions (dinatural transformations) between sets $H\ R\ R$ and
$\mathcal{C}(F~R~R, C)$.

We translate this definition to Cedille, where
an $H$-restricted $F$-cowedge $(C, \Phi)$ corresponds to a type
(\verb;C;) and a polymorphic function (\verb;RCowedge H F C;):
\begin{alltt}
RCowedge ◂ (★ ➔ ★) ➔ (★ ➔ ★) ➔ ★ ➔ ★ 
 = λ H. λ F. λ C. ∀ R : ★. H · R ➾ F · R ➔ C.
\end{alltt}
To simplify the subsequent development, we render difunctors as schemes
with a single parameter, and the restriction \verb;H R; is made
implicit (denoted by  \verb;➾;, which is a non-dependent version of
\verb;∀;). The simplification of making the restriction parameter
erased allows us to avoid needing function extensionally to achieve
our encoding.
The carrier of the initial cowedge can be implemented in terms of
universal quantification:
\begin{alltt}
RCoend ◂ (★ ➔ ★) ➔ (★ ➔ ★) ➔ ★ = λ H. λ F. ∀ C : ★. RCowedge · H · F · C ➔ C.
\end{alltt}
The second component of initial cowedges is a polymorphic function,
(\verb;intrRCoend;), which plays the role of the constructor of its carrier
(\verb;RCoend H F;), and is implemented as follows:
\begin{alltt}
intrRCoend ◂ ∀ H F : ★ ➔ ★. RCowedge · H · F (RCoend · H · F)
  = Λ H. Λ F. Λ R. Λ ac. λ ga. (Λ Y. λ q. q · R -ac ga).
\end{alltt}
The (weak) initiality can be proved by showing that for any cowedge
\verb;RCowedge H F C;, there is a homomorphism from \verb;RCoend H F; to
\verb;C;:
\begin{alltt}
elimRCoend ◂ ∀ H F: ★ ➔ ★.∀ C: ★. RCowedge · H · F · C ➔ RCoend · H · F ➔ C 
 = Λ F. Λ A. Λ C. λ phi. λ e. e phi.
\end{alltt}

\subsection{Dependent Elimination for Restricted Coends}
In this section, we utilize the intersection type (denoted by \verb;ι;
in Cedille) to define a restricted coend type for which the induction
principle is provable.  To do this, we follow the original recipe
described by Stump to derive natural-number induction in
Cedille. First, we define a predicate expressing that an
\verb;H;-restricted \verb;F;-coend is inductive.
\begin{alltt}
RCoendInductive ◂ Π H F : ★ ➔ ★. RCoend · H · F ➔ ★
  = λ H. λ F. λ e. ∀ Q : RCoend · H · F  ➔ ★. 
  (∀ R : ★. ∀ hr : H · R. Π fr : F · R. Q (intrRCoend -hr fr)) ➔ Q e.
\end{alltt}
Second, we define the ``true'' inductive restricted coend as an
intersection of the previously defined \verb;RCoend; and the predicate
\verb;RCoendInductive;. In essence, this says that \verb;RCoendInd; is
the subset of \verb;RCoend; carved out by the \verb;RCoendInductive;
predicate.
\begin{alltt}
RCoendInd ◂ (★ ➔ ★) ➔ (★ ➔ ★) ➔ ★
 = λ H. λ F. ι x : RCoend · H · F. RCoendInductive · H · F x.
\end{alltt}
This definition builds on an observation by Leivant that under the
Curry-Howard isomorphism, proofs in second-order logic that data
satisfy their type laws can be seen as isomorphic to the
Church-encodings of those data~\cite{leivant83}.
Next, we define the constructor for the inductive coend:
\begin{alltt}
intrRCoendInd ◂ ∀ H F : ★ ➔ ★. RCowedge · H · F (RCoendInd · H · F) 
 = Λ H. Λ F. Λ R. Λ hr. λ fr. 
 [ intrRCoend -hr fr , Λ Q. λ q. q · R -hr fr ].
\end{alltt}
In Cedille, the term \verb;[ t , t' ]; introduces the intersection type
\verb;ι x : T. T' x;, where \verb;t; has type \verb;T; and \verb;t';
has type \verb;[t/x]T';. Definitionally, values of intersection types
reduce (via erasure) to their first components (i.e.,
\verb;[ t , t' ]; is definitionally equal to \verb;t;). See~\cite{firsov18b} for
more information on intersection types in Cedille.
The induction principle is now derivable and has the following type:
\begin{alltt}
indRCoend ◂ ∀ H F : ★ ➔ ★. ∀ Q : RCoendInd · H · F ➔ ★.
  (∀ R : ★. ∀ hr : H · R. Π fr : F · R. Q (intrRCoendInd  -hr fr))
   Π e : RCoendInd · H · F. Q e = <..>
\end{alltt}

\subsection{Identity Restricted Existentials}
We define the identity restricted existential of \verb;F; and the object
\verb;C; as an \verb;F;-coend restricted by a family of identity
functions \verb;λ X. Id · X · C;:
\begin{alltt}
RExtInd ◂ (★ ➔ ★) ➔ ★ ➔ ★ = λ F. λ X. RCoendInd · (λ R : ★. Id R X) · F.
\end{alltt}
Next, we prove that the restricted existential of any \verb;F; is
an identity mapping:
\begin{alltt}
imapRExt ◂ ∀ F : ★ ➔ ★. IdMapping · (RExtInd · F) 
 = Λ F. Λ A. Λ B. Λ f. λ c. indRCoend c
   (Λ R. Λ i. λ gr. pair (intrRExtInd -(compose i f) gr) β).
\end{alltt}
Intuitively, \verb;RExtInd F X; corresponds to the type
\verb;∃ R. Id R X × F R;.
Notice that \verb;RExtInd F X; is positive
because \verb;X; occurs positively in \verb;Id;, and that positivity
does not depend on \verb;F;.  With the definition of identity
restricted existentials in place, we can now move on towards using
them to derive course-of-value induction.

\section{Course-of-Value Datatypes}
\label{course-of-value-datatypes}
In this section we review why the naive scheme for an inductive
datatype with a destructor does not work out due to negativity (as
mentioned in the introduction, Section~\ref{intro}). Then, we demonstrate
how \textit{identity restricted existentials}
(\verb;RExtInd; of Section~\ref{restricted-existentials})
can be used in our novel
encoding to overcome this limitation, allowing us to define datatypes
supporting course-of-value induction.
The development in this section is parameterized by an identity
mapping:
\begin{alltt}
module (F : ★ ➔ ★)\{imap : IdMapping · F\}.
\end{alltt}

\subsection{Precursor}
\label{precursor}
In~\cite{uustalu99}, Uustalu and Vene showed that it is possible to
use restricted existentials to derive a superset of course-of-value
natural numbers. We start by generalizing their construction to
arbitrary inductive types, in terms of least fixed points of identity
mappings.
The main idea is to define a combinator that pairs the value
\verb;F X; with the destructor function (of type \verb;X ➔ F X;):
\begin{alltt}
Enr' ◂ ★ ➔ ★ = λ X. F X × (X ➔ F X).
\end{alltt}
Intuitively, we wish to construct a least fixed point of \verb;F; and
its destructor simultaneously. The resulting scheme \verb;Enr' F; is
not positive and therefore it cannot be a functor nor an identity
mapping. This implies that we cannot take a least fixed point of it
directly.  Instead, we define \verb;CVF' · F; as a restricted
existential of \verb;Enr' F;. Hence, the scheme \verb;CVF' F; is an
identity mapping by the property of restricted existentials:
\begin{alltt}
CVF' ◂ ★ ➔ ★ = RExtInd · (Enr' · F).
imCVF' ◂ IdMapping (CVF' F) = imapRExt · (Enr' · F).
\end{alltt}
It is natural to ask what the relationship between the least fixed
point of \verb;F; and least fixed point of \verb;CVF' F; is.
\begin{alltt}
FixCV' ◂ ★ = FixIndM · (CVF' · F) -(imCVF' · F).
\end{alltt}
It turns out that \verb;FixCV'; is not a least fixed point of
\verb;F;, because value \verb;F FixCV'; could be paired with any
function of type \verb;FixCV' ➔ F FixCV';.  We will provide more
intuition by describing the destructor and constructor functions of
\verb;FixCV';.

\paragraph{Destructor} The generic development from Section~\ref{inductive-datatypes} allows us to unroll
\verb;FixCV'; into a value of \verb;CVF' FixCV'; (which it was made
from). Because \verb;CVF' F; is a restricted existential, we can use
its dependent elimination to ``project out'' the value
\verb;F FixCV';:
\begin{alltt}
outCV' ◂ FixCV' ➔ F · FixCV' = λ x. indRExt (outFixIndM -imapRExt x)
   (Λ R. Λ c. λ v. elimId -(imap -c) (π₁ v)).
\end{alltt}
In the definition above, the variable \verb;v; has type
\verb;F R × (R → F R);. Because \verb;F; is an identity mapping, we
can cast the first projection of \verb;v; to \verb;F FixCV'; and return
it. On the other hand, the function \verb;R → F R; cannot be casted to
type \verb;FixCV' ➔ F FixCV';, because the abstract type \verb;R; appears
both positively and negatively.

\paragraph{Constructor} Similarly, the generic development gives us the function
\verb;inFixIndM;, which constructs a \verb;FixCV'; value from
the given \verb;CVF' FixCV';. The latter must be built
from a pair of \verb;F FixCV'; and a function of type
\verb;FixCV' ➔ F FixCV';. This observation gives rise
to the following specialized constructor of \verb;FixCV';:
\begin{alltt}
inCV' ◂ (FixCV' ➔ F FixCV') ➔ F FixCV' ➔ FixCV' = <..>
\end{alltt}
This constructor indicates that \verb;FixCV'; represents the superset
of course-of-value datatypes, because the function
\verb;FixCV' → F FixCV'; is not restricted to the destructor
\verb;outCV';, and the inductive value might contain a different
function of that type at every construtor. We address this issue in
the next section.

\subsection{Course-of-Value Datatypes with Induction}
\label{cov-induction}
In our previous work \cite{firsov18b}, we developed a generic unrolling function 
for least fixed points of identity mappings:
\begin{alltt}
outFixIndM ◂ ∀ imap : IdMapping F. FixIndM F ➔ F (FixIndM F) = <..>
\end{alltt}
Observe that the only identity-mapping-specific variable is quantified
implicitly.  In other words, \verb;outFixIndM; does not perform any
\verb;F;-specific computations. The same is true for the elimination
principle of restricted existentials (\verb;indRCoend;). Since \verb;outCV'; is implemented
in terms of these functions, this observation
suggests that we can refer to \verb;outCV'; as we define the
subset of type \verb;FixCV';.  In particular, we define the scheme
\verb;Enr; by pairing the value \verb;F X; with the function
\verb;f : X ➔ F X; and the proof that this function is equal to the
previously defined \verb;outCV';:
\begin{alltt}
Enr ◂ ★ ➔ ★ = λ X. F · X × Σ f : X ➔ F · X. f ≃ outCV'.
\end{alltt}
This constraint between terms of different types is possible due to
heterogeneous equality. Just like in the previous section, we define a
least fixed point of the restricted existential of \verb;Enr F; and
its least fixed point:
\begin{alltt}
CVF ◂ ★ ➔ ★ = λ X. RExtInd · (Enr · F) X.
FixCV ◂ ★ = FixIndM · CVF (imapRExt · Enr).
\end{alltt}
\paragraph{Destructor}
The destructor of \verb;FixCV; is represented by exactly the same
lambda-term as the destructor (\verb;outCV';) of \verb;FixCV';:
\begin{alltt}
outCV ◂ FixCV ➔ F · FixCV = λ v. indRExt 
 (outFixIndM  -imapRExt v) (Λ R. Λ c. λ v. elimId -(imap -c) (π₁ v)).
outCVEq ◂ outCV' ≃ outCV = β.
\end{alltt}
Because the only difference between \verb;outCV'; and
\verb;outCV; is their typing annotations (which are inferred by
the typechecker), they are definitionally equal in Cedille
(as witnessed by \verb;β;, the introduction rule of Cedille's
equality type).
\paragraph{Constructor}
Armed with the destructor \verb;outCV; and the proof \verb;outCVEq;,
we can now define the constructor of \verb;FixCV;:
\begin{alltt}
inCV ◂ F · FixCV ➔ FixCV = Λ F. Λ imap. λ fcv. inFixIndM -imapRExt
 (intrRExtInd -trivIdExt) (pair fcv (pair (outCV -imap) outCVEq)).
\end{alltt}
\paragraph{Lambek's Lemma}
As expected, \verb;inCV; and \verb;outCV; are mutual
inverses, which establishes that \verb;FixCV; is a fixed point of
\verb;F;.
\begin{alltt}
lambekCV1 ◂ ∀ x : F FixCV. outCV (inCV x) ≃ x = β.
lambekCV2 ◂ ∀ x : FixCV.   inCV (outCV x) ≃ x = <..>
\end{alltt}

Note that \verb;lambekCV1; holds definitionally, while \verb;lambekCV2;
is provable by straightforward induction (actually, the proof only
uses dependent case analysis, and ignores the inductive hypothesis).

\paragraph{Induction}
Recall that the induction principle for the least fixed point \verb;FixIndM;
is stated in terms of proof-algebras
(\verb;PrfAlgM; in Section~\ref{inductive-datatypes}). Now, let us define
proof-algebras for course-of-value datatypes:
\begin{alltt}
PrfAlgCV ◂ (FixCV ➔ ★) ➔ ★ = ∀ R : ★. ∀ c : Id · R · FixCV. 
 Π out : R ➔ F · R. out ≃ outCV ➾ Π ih : Π r : R. Q (elimId -c r).
 Π fr : F · R.  Q (inCV (elimId -(imap -c) fr))
\end{alltt}
Compared to \verb;PrfAlgM;~\cite{firsov18b},
\verb;PrfAlgCV; has an extra argument (\verb;out : R ➔ F R;),
which represents an unrolling (or, abstract predecessor) function for
abstract type \verb;R;. Additionally, we have a proof that the \verb;out; function
is equal to the previously discussed destructor \verb;outCV;. This evidence
is needed when the construction of a particular
proof-algebra depends on the exact definition of the predecessor
function. Besides those two extra arguments,
the identity function (\verb;c;) from abstract \verb;R; to concrete
\verb;FixCV;, the inductive hypothesis (\verb;ih;), and the subdata
(\verb;fr;) are the same as in \verb;PrfAlgM;~\cite{firsov18b}.

Course-of-value induction is expressible in terms of course-of-value
proof-algebras and is proved by combining the induction principle of
\verb;FixIndM; with the dependent elimination principle of restricted existentials.
\begin{alltt}
inductionCV ◂ ∀ Q : FixCV ➔ ★. PrfAlgCV Q ➔ Π x : FixCV. Q x = <..>
\end{alltt}
It is important to establish the computational behaviour of this proof-principle:
\begin{alltt}
indCancel ◂ ∀ Q : FixCV ➔ ★. ∀ palg : PrfAlgCV Q. ∀ x : F FixCV.
 inductionCV palg (inCV x) ≃ palg outCV (inductionCV palg) x = β.
\end{alltt}
Above, notice how the abstract unrolling function \verb;out : R ➔ F R; is being
instantiated with the actual unrolling function \verb;outCV;.
Finally, implementing course-of-value recursion (\verb;foldCV;) from
Section~\ref{intro} in terms of course-of-value induction
(\verb;inductionCV;) is straightforward.

\section{Examples}
\label{examples}
We now demonstrate the utility of our results with example functions and
proofs on natural numbers that require course-of-value recursion and induction. Note that the
\verb;fibCV; example from the introduction (Section~\ref{intro})
works as described, because \verb;foldCV; is derivable from \verb;inductionCV;.
Recall that natural numbers may be defined as the least fixed point of a functor
(\verb;NF ◂ ★ ➔ ★ = λ X. Unit + X.;).
As remarked in Section~\ref{identity-mapping}, because \verb;NF; is a functor, it is
also an identity mapping (\verb;nfimap ◂ IdMapping NF = <..>;).
We begin by defining the type of natural numbers (\verb;NatCV;), supporting
a constant-time predecessor function, as well as
course-of-value induction:
\begin{alltt}
NatCV ◂ ★ = FixCV F nfimap.
zero ◂ NatCV = inCV -nfimap (in1 unit).
suc ◂ NatCV ➔ NatCV = λ n. inCV -nfimap (in2 n).
pred ◂ NatCV ➔ NatCV = λ n. case (outCV -nfimap n) (λ u. n) (λ n'. n').
\end{alltt}

\subsection{Division}
\label{examples:div}
Consider an intuitive definition of division as iterated subtraction:
\begin{alltt}
div : Nat ➔ Nat ➔ Nat
div 0 m = 0
div n m = if (n < m) then 0 else (suc (div (n - m) m))
\end{alltt}
Such a definition is rejected by Agda (and many languages like it),
because Agda requires that recursive calls are made on arguments its
termination checker can guarantee are structurally smaller, which it
cannot do for an arbitrary expression (like \verb;n - m;). With our
development, the problematic recursive call (on \verb;n - m;) is an
instance of course-of-value recursion because we can define subtraction
by iterating the predecessor function, and we have access to recursive
results for every predecessor.

For convenience, we define the conventional \verb;foldNat; as a
specialized version of our generic development. Then, \verb;minus n m;
is definable as the \verb;m; number of predecessors of \verb;n;.
\begin{alltt}
foldNat ◂ ∀ R : ★. (R ➔ R) ➔ R ➔ NatCV ➔ R
 = Λ R. λ rstep. λ rbase. foldCV (Λ R'. λ out. λ rec. λ nf. 
 case nf (λ _. rbase) (λ r'. rstep (rec r'))).
minus' ◂ ∀ R : ★. (R ➔ NF R) ➔ R ➔ NatCV ➔ R
  = Λ R. λ pr. foldNat (λ r. case (pr r) (λ _. r) (λ r'. r'))
minus ◂ NatCV ➔ NatCV ➔ NatCV = minus' (outCV -nfimap).
\end{alltt}

Above, we first define an abstract operation \verb;minus' n m;, where
the type of \verb;n; is polymorphic and where that type comes with an
abstract predecessor \verb;pr;. Then, the usual concrete
\verb;minus n m; is recovered by using \verb;NatCV; for the
polymorphic type and the destructor \verb;outCV -nfimap; for the
predecessor.

Now we can use \verb;minus'; to define division naturally, returning
zero in the base case, and iterating subtraction in the step
case. This definition below is accepted purely through type-checking and
without any machinery for termination-checking.

\begin{alltt}
div ◂ NatCV ➔ NatCV ➔ NatCV
  = λ n. λ m. inductionCV (Λ R. Λ c. λ pr. Λ preq. λ ih. λ nf.
  case nf
    (λ x. zero)                           % div 0 m
    (λ r. if (suc (elimId -c r) < m)      % div (suc n) m
      then zero
      else (suc (ih (minus' pr r (pred m)))))) n
\end{alltt}

Notice that in the conditional statement, we use \verb;elimId -c; to
convert the abstract predecessor \verb;r; to a concrete natural number,
allowing us to apply \verb;suc; to check if \verb;suc n; is less than \verb;m;.
In the intuitive definition of \verb;div;, we match on \verb;0; in
the first case, and on any wildcard pattern \verb;n; in
the second case. In contrast, when using \verb;inductionCV; and \verb;case;
in our example, we must explicitly handle the \verb;zero; and \verb;suc r;
cases. Consequently, while the intuitive definition recurses on
\verb;div (n - m) m;, we recurse on the predecessors
\verb;ih (minus' pr r (pred m));. This is equivalent because
\verb;minus (suc n) (suc m); is equal to \verb;minus n m;, for all numbers
\verb;n; and \verb;m;. We can also prove properties about our
development, such as the aforementioned equivalence
(\verb;minSucSuc; below, proven by ordinary induction). By direct
consequence, we can also prove that the defining equation
(\verb;divSucSuc; below, for the
successor case) of the intuitive definition of division holds (by
ordinary induction and rewriting by \verb;minSucSuc;):
\begin{alltt}
minSucSuc ◂ Π n m : NatCV. minus (suc n) (suc m) ≃ minus n m = <..>
divSucSuc ◂ Π n m : NatCV. (suc n < suc m) ≃ ff ➔
  div (suc n) (suc m) ≃ suc (div (minus (suc n) (suc m)) (suc m)) = <..>
\end{alltt}
While the propositions are stated in terms of concrete
\verb;minus;, the \verb;div; function is defined in terms of abstract
\verb;minus';. Nonetheless, the propositions are provable due to the
computational behavior of \verb;inductionCV;, which instantiates
the bound \verb;pr; with \verb;outCV;, allowing us to identify
\verb;minus' pr; and \verb;minus;.

\subsection{Property of Division}
\label{examples:divprop}

Besides this section, this paper contains (mostly omitted) proofs by
definitional equality, rewriting, dependent
case-analysis, and/or ordinary induction.
In this section we prove a property of division that takes full
advantage of course-of-value induction (our primary contribution).

We assume the existence of a less-than-or-equal relation
(whose definition is omitted for space reasons)
on course-of-value naturals (\texttt{LE ◂ NatCV ➔ NatCV ➔ ★}), with
two constructors for evidence in the zero case
(\texttt{leZ ◂ Π n : NatCV. LE zero n}),
and in the successor case
(\texttt{leS ◂ ∀ n m : NatCV. LE n m ➔ LE (suc n) (suc m)}).
Additionally, we will need a lemma that \verb;LE; is transitive
(\texttt{leTrans ◂ ∀ x y z : NatCV. LE x y ➔ LE y z ➔ LE x z}),
and a lemma that subtraction decreases a number or keeps it the same
(\texttt{leMinus ◂ Π n m : NatCV. LE (minus n m) n}). Both of these
lemmas are provable by ordinary induction.
Now, let's prove our property of interest by
\textit{course-of-value induction}, namely that division also
decreases a number or keeps it the same:\footnote{To make the proof
  easier to read, we use non-depenent \texttt{if} and \texttt{case}
  statements. Our actual code requires the dependent eliminator
  counterparts of these statements, along with the appropriate motives.}
\begin{alltt}
leDiv ◂ Π n m : NatCV. LE (div n m) n
  = λ n. λ m. inductionCV (Λ R. Λ c. λ pr. Λ preq. λ ih. λ nf.
  case nf
    (λ u. ρ (etaUnit u) - leZ zero)  % Goal: LE (div zero m) zero
    (λ r. let n1 = elimId -c r in    % Goal: LE (div (suc n) m) (suc n)
     if (suc n1 < m)
     then (leZ (suc n1)) % Goal: LE zero (suc n)
     else let            % Goal: LE (suc (div (minus n (pred m)) m)) (suc n)
       n2 = minus n1 (pred m)
       n3 = div n2 m
       n3LEn2 = ρ (sym preq) - ih (minus' pr r (pred m))
       n2LEn1 = leMinus n1 (pred m)
       in leS -n3 -n1 (leTrans -n3 -n2 -n1 n3LEn2 n2LEn1))).
\end{alltt}

\paragraph{Zero Case} 

The expression \verb;div zero m; in the goal reduces to \verb;zero;,
allowing us to conclude \verb;LE zero zero; by using the constructor
\verb;leZ;. The only caveat is that the reduction requires rewriting
(using Cedille's \verb;ρ; primitive) by the uniqueness
of unit (\verb;etaUnit;), because our generic encoding of \verb;NatCV;
uses the unit type in the left part of the sum (\verb;NF X = 1 + X;).

\paragraph{Successor Case (When the Conditional is True)}

At first, the successor case of division is prevented from reducing
further because it is branching on a conditional statement
(\texttt{suc (elimId -c r) < m}). In the true branch of this
conditional, the goal reduces and is immediately solvable using \verb;leZ;.
Note that we use a let statement to name the result
(variable \verb;n1;) of converting the
abstract predecessor \verb;r : R; to a concrete \verb;NatCV; via the
expression \verb;elimId -c r;.

\paragraph{Successor Case (When the Conditional is False)}

The goal in the false branch of the conditional is solvable by using
\verb;leS;, leaving us with the subgoal
\texttt{LE (div (minus n (pred m)) m) (suc n)}. We name the inner
subtraction-expression \verb;n2;, and the outer division-expression
\verb;n3;. We solve the subgoal (\verb;LE n3 n1;)
using transitivity (\verb;leTrans;) by showing that
\verb;n3; $\leq$ \verb;n2; $\leq$ \verb;n1;, leaving us with two final
subsubgoals. The second subsubgoal (\verb;LE n2 n1;) is provable
(\verb;n2LEn1;) using our lemma about subtraction getting smaller or
staying the same (\verb;leMinus;).

Finally, the first subsubgoal (\verb;LE n3 n2;) is the interesting
case (\verb;n3LEn2;). First, we rewrite using \verb;preq; to change
our goal from requiring the concrete natural number predecessor
(\verb;outCV -nfimap;) to instead require the abstract predecessor
(\verb;pr;). We can use \verb;ih; to get an inductive hypothesis,
of type \verb;LE (div (elimId -c r) m) (elimId -c r);, for any
\textit{abstract} natural number \verb;r;. Thus, our final subsubgoal is
solvable by the inductive hypothesis where \verb;r; is the result of
the abstract subtraction \verb;minus' pr r (pred m);. Because
\verb;elimId -c; is definitionally equal to the identity function, and
because we already rewrote by \verb;preq;, this gives us exactly what
we want. This is despite the fact that our original subsubgoal
(\verb;LE n3 n2;) is stated in terms of \verb;n3; and \verb;n2;, which
use concrete \verb;div; and \verb;minus;, respectively!
Importantly, the inductive hypothesis we use requires course-of-value
induction, obtained by an expression that iterates the
predecessor function (\verb;ih (minus' pr r (pred m));).
In contrast, ordinary induction corresponds to using the inductive
hypothesis \verb;ih r;.

\subsection{Catalan Numbers}
Many solutions to counting problems in combinatorics can be given in terms of
Catalan numbers. The Catalan numbers are definable as the solution to
the recurrence $C_0 = 1$ and $C_{n + 1} = \sum_{i=0}^n C_i C_{n - i}$.
This translates to an intuitive functional definition of
the Catalan numbers:
\begin{alltt}
cat : Nat → Nat
cat 0 = 1
cat (suc n) = sum (λ i → cat i * cat (n - i)) n
\end{alltt}
The \verb;sum; function has type \verb;(Nat ➔ Nat) ➔ Nat ➔ Nat;, where
the lower bound of the sum is always zero (\verb;i=0;), the second argument is the
upper bound of the sum (\verb;n;), and the first argument is the body of the sum
(parameterized by \verb;i;).
Once again, this is not a structurally terminating function
recognizable by Agda. While \verb;fib; and \verb;div; have a static
number of course-of-value recursions (two and one, respectively), the
number of recursions made by \verb;cat; is determined by its
input. Nonetheless, we are able to define \verb;cat; using our development.
\begin{alltt}
cat ◂ NatCV ➔ NatCV
  = inductionCV (Λ R. Λ c. λ pf. Λ pfeq. λ ih . λ nf.
  case nf
    (λ _. suc zero)  % cat 0
    (λ r. sum        % cat (suc n)
      (λ i. mult
        (ih (minus' pf r (minus (elimId -c r) i)))
        (ih (minus' pf r i)))
      (elimId -c r))).
\end{alltt}
As with \verb;div;, above \verb;r; has abstract type \verb;R;, so we
convert it to a \verb;NatCV; where necessary by applying
\verb;elimId -c;. The intuitive right factor \verb;cat (n - i); is
directly encoded as \verb;ih (minus' pf r i));. However, we cannot
directly encode the intuitive left factor \verb;cat i;, because
\verb;i; is a natural number and we only have inductive hypotheses for
values of abstract type \verb;R;. However, \verb;i; is equivalent to
\verb;n - (n - i); for all \verb;i; where \verb;i ≤ n;.
We use the abstract \verb;minus';
function for the outer subtraction, whose first numeric argument is an
abstract \verb;r; but whose second numeric argument expects a concrete
\verb;NatCV;. Hence, the inner subtraction is a concrete \verb;minus;,
whose first argument is the concrete version of \verb;r; (converted
via identity function \verb;c;) and whose second argument is the
concrete \verb;i; (of type \verb;NatCV;). Because the outer
subtraction (\verb;minus';) returns an abstract R, we can get an
inductive hypothesis for an expression equivalent to \verb;i;.
Finally, we can prove the
aforementioned equivalence for \verb;minus; (by rewriting), and as a consequence the
defining equation (for the successor case) of the intuitive definition
of Catalan numbers (by dependent case-analysis and rewriting by \verb;minusId;):
\begin{alltt}
minusId ◂ Π n i : NatCV. (i ≤ n) ≃ tt ➔ minus n (minus n i) ≃ i = <..>
catSuc ◂ Π n : NatCV.
  cat (suc n) ≃ sum (λ i. mult (cat i) (cat (minus n i))) n = <..>
\end{alltt}
Once again, the discrepancy between abstract \verb;minus'; and
concrete \verb;minus; is resolved in the proofs thanks to the computational
behavior of \verb;inductionCV; instantiating \verb;pr; to \verb;outCV;.
      
\section{Conclusions and Related Work}
\label{related-work}

Ahn et al.~\cite{ahn11} describe a hierarchy of Mendler-style recursion
combinators. They implement generic course-of-value
recursion in terms of Haskell's general recursion. Then, they
prove that course-of-value recursion for arbitrary ``negative''
inductive datatypes implies non-termination.

Miranda-Perea~\cite{MIRANDAPEREA2009103} describes extensions of System~F with primitive
course-of-value iteration schemes. He
explains that the resulting systems lose strong normalization if they
are combined with negative datatypes.

Uustalu et al.~\cite{UUSTALU2002315} show that natural-deduction proof
systems for intuitionistic logics can be safely \textit{extended} with
a course-of-value induction operator in a proof-theoretically
defensible way. Their requirement of monotonicity corresponds to our
notion of \verb;IdMapping;.

In contrast to the work described above, we have now shown that
course-of-value induction can be \textit{derived} within type theory
(specifically, within CDLE).  We will end with comparing our approach
to alternative approaches to handling complex termination arguments
within type theory. \footnote{
  For comparison, our accompanying code includes Agda formalizations of \texttt{fib},
  \texttt{div}, and \texttt{cat} in the ``Below'' style of
  Section~\ref{related-work:below} and the sized types style of
  Section~\ref{related-work:sized}.
}

\subsection{The Below Way}
\label{related-work:below}

Goguen et al.~\cite{goguen06} define the induction principle
\verb;recNat; (generalizing to all inductive types), which they use to
elaborate dependent pattern matching to eliminators. In the step
case, \verb;recNat; receives \verb;BelowNat P n;, which is a large
tuple consisting of the motive \verb;P; for every predecessor of
\verb;n;. Simple functions performing nested pattern matching
(e.g., \verb;fib;) can be written using \verb;recNat; and
nested case-analysis, by projecting out inductive hypotheses from
\verb;BelowNat P n;. However, functions with more complex termination
arguments (e.g., \verb;div; and \verb;cat;) require proving extra
lemmas (e.g., \verb;recMinus; in our accompanying code) to dynamically
extract inductive hypotheses from \verb;BelowNat P n; evidence. In our
approach, such lemmas are unnecessary.

\subsection{Sized Types}
\label{related-work:sized}

Abel~\cite{abel-sized-types} extends type theory with a notion of
\textit{sized types}, which allows intuitive function
definitions to be accepted by termination checking.  Course-of-value
induction (CoVI) and sized types (ST) have trade-offs. ST requires
defining size-indexed versions of the datatypes, which necessitates
altering conventional type signatures of functions to include size
information. While CoVI is
derivable within CDLE, ST extends the underlying type
theory. On the other hand, CoVI is restricted to functions that
recurse strictly on previous values. Hence, a function like merge sort
can be written using ST but not with CoVI. As future work, we would
like to investigate datatype encodings with a restricted version of
abstract constructors (in addition to abstract destructors) for
defining functions like merge sort.



\bibliography{paper}

\end{document}